\pdfoutput=1 
\documentclass[
aip, 
pccp, 
reprint,
floatfix,
amsmath,amssymb,]{revtex4-1}

\usepackage{graphicx}
\usepackage[utf8]{inputenc}
\usepackage[T1]{fontenc}
\usepackage{mathptmx}
\usepackage{etoolbox}
\usepackage{hyperref}
\usepackage{bm}

\usepackage{natbib}
\setcitestyle{square, comma, numbers,sort&compress, super}

\usepackage[version=3]{mhchem}
\usepackage[left=1.5cm, right=1.5cm, top=1.785cm, bottom=2.0cm]{geometry}
\usepackage{balance}
\usepackage{lastpage}
\usepackage[format=plain,justification=justified,singlelinecheck=false,font={stretch=1.125,small,sf},labelfont=bf,labelsep=space]{caption}

\usepackage{float}
\usepackage{fancyhdr}
\usepackage{fnpos}
\usepackage[english]{babel}
\addto{\captionsenglish}{%
  
}
\usepackage{array}
\usepackage{droidsans}
\usepackage{charter}

\usepackage[usenames,dvipsnames]{xcolor}
\usepackage{setspace}
\usepackage[compact]{titlesec}

\usepackage{booktabs}
\usepackage{dcolumn}
\usepackage{chemfig}
\usepackage{dcolumn}

\usepackage{epstopdf}

\definecolor{cream}{RGB}{222,217,201}

\usepackage{amsmath}
\usepackage{amssymb}

\def\cm{cm$^{-1}$}

\def\ie{\textit{i.e.}}
\def\eg{\textit{e.g.}}

\def\pesfinal{FAD-HDNNP}
\newcommand{\pesdevel}[1]{HDNNP#1}
\def\abinit{\textit{ab initio}}
\newcommand{\com}[1]{\textsuperscript{\emph{#1}}}

\draft 

\makeatletter
\def\@email#1#2{%
 \endgroup
 \patchcmd{\titleblock@produce}
  {\frontmatter@RRAPformat}
  {\frontmatter@RRAPformat{\produce@RRAP{*#1\href{mailto:#2}{#2}}}\frontmatter@RRAPformat}
  {}{}
}%
\makeatother

\begin{document}

\title{High-dimensional neural network potentials for accurate vibrational frequencies: The formic acid dimer benchmark}

\author{Dilshana Shanavas Rasheeda$^{{\ast}1}$,Alberto Mart\'in Santa Dar\'ia\textit{$^{2}$}, Benjamin Schr\"oder$^{1}$,Edit M\'atyus$^{2}$, J\"org Behler}

\affiliation{\textit{$^{1}$~Universit\"at G\"ottingen, Institut f\"ur Physikalische Chemie, G\"ottingen, Germany
\\
$^{2}$~Eötvös Loránd University, Institute of Chemistry, Budapest, Hungary}}
\email{dilshana.rasheeda@chemie.uni-goettingen.de}

\date{\today}

\begin{abstract}
In recent years, machine learning potentials (MLP) for atomistic simulations have attracted a lot of attention in chemistry and materials science. Many new approaches have been developed with the primary aim to transfer the accuracy of electronic structure calculations to large condensed systems containing thousands of atoms. 

In spite of these advances, the reliability of modern MLPs in reproducing the subtle details of the multi-dimensional potential-energy surface is still difficult to assess for such systems. On the other hand, moderately sized systems enabling the application of tools for thorough and systematic quality-control are nowadays rarely investigated.

In this work we use benchmark-quality harmonic and anharmonic vibrational frequencies as a sensitive probe for the validation of high-dimensional neural network potentials. For the case of the formic acid dimer, a frequently studied model system for which stringent spectroscopic data became recently available, we show that high-quality frequencies can be obtained from state-of-the-art calculations in excellent agreement with coupled cluster theory and experimental data.
\end{abstract}


\maketitle 

\section{Introduction}\label{sec:introduction}

Due to its unique capabilities to process and analyze large amounts of data, machine learning has nowadays found numerous applications in chemistry and related fields~\cite{P5590,P5779,RNN1993,RNN2-2018,P6328}. Starting with the work of Doren and coworkers in 1995~\cite{P0316}, a prominent example is the representation of the atomic interactions with quantum mechanical accuracy by learning the potential energy surface (PES) from a set of known training points computed using accurate electronic structure methods. The resulting machine learning potentials (MLP) have rapidly evolved in the past two decades~\cite{P4885,P5673,P5793,P6102,P6112,P6121}. While first-generation MLPs have been restricted to small molecular systems \cite{P2559,P3033}, second-generation MLPs like high-dimensional neural network potentials~\cite{P1174}, Gaussian approximation potentials~\cite{P2630}, spectral neighbor analysis potentials~\cite{P4644}, moment tensor potentials~\cite{P4862}, atomic cluster expansion~\cite{P5794} and many others have paved the way to large-scale simulations of condensed systems, from water and aqueous solutions to bulk materials and interfaces. Long-range electrostatic interactions~\cite{P2962,P5313,P5577}, non-local charge transfer~\cite{P4419,P5932,P5859,P6122,P5977}, and magnetism~\cite{P6057,P6000} can also be explicitly considered in modern MLPs.

Due to their very flexible but simple functional form MLPs offer advantages like a very high accuracy of about 1~meV/atom in reproducing known reference total energies, an efficiency close to empirical force fields, and an unbiased description of many types of atomic interactions -- from covalent bonds via dispersion interactions to metallic and ionic bonding. Still, although some recent MLPs include physical terms like electrostatics, in general the high flexibility of MLPs is curse and bless at the same time, since the functional forms employed in MLPs, like neural networks or kernel methods, are not guaranteed to yield the correct physical shape of the PES. Therefore, large training sets are needed to ensure that a reliable representation of the PES is obtained in the training process. Moreover, a careful validation is required, since energy and force predictions can be inaccurate when extrapolating to atomic configurations that are too different from those in the training sets.

These limited extrapolation capabilities of MLPs and the need for a dense enough sampling of the reference data set raise the question how modern MLPs can be validated. While for early MLPs constructed for small molecular systems a systematic, grid-based mapping of the underlying PES has been possible allowing for rigorous quality control, the situation is different for second-generation MLPs designed for very large systems. Here, the total energy is typically constructed as a sum of environment-dependent atomic energies, resulting in a linear scaling of the computational effort with system size enabling simulations of thousands of atoms. However, since the atomic energies, which are not quantum mechanical observables, can be considered as mathematical auxiliary quantities, error compensation might occur reducing the transferability of the potentials. Further, typically investigated quantities like the root mean squared errors (RMSE) of energies and forces are of limited use, since they can only be computed for the available reference structures, and as averaged properties they may not allow to assess the quality of all the fine details of the PES.

Here, we propose to employ the harmonic and anharmonic vibrational frequencies as a sensitive probe to validate the quality of MLPs pinpointing on close-to equilibrium structures, which are particularly important for spectroscopic applications. While far-from equilibrium behaviour is in principle equally interesting also for spectroscopy, any PES not being able to describe the equilibrium properties correctly could not be trusted for spectroscopic applications.
This tool can thus be viewed as complementary to other procedures like active learning that are covering a wide range of non-equilibrium structures~\cite{P3114,P4939,P5399}. 
Vibrational frequencies have been studied since the advent of MLPs, mainly for small molecular systems, with great success~\cite{P0829,P0840}, and very accurate approaches for constructing MLPs suitable for this purpose have been derived~\cite{P1390,P2210,P5213}. Still, most of these methods exhibit an unfavorable scaling with system size. Second-generation potentials allow to address larger systems, and some vibrational studies for larger molecules~\cite{P5205}, cluster~\cite{P3132} and condensed systems~\cite{P5600,P5570,P6134,P6196} have been reported. However, MLPs suitable for condensed systems have been typically constructed relying on density functional theory (DFT), which, although offering a good compromise between accuracy and efficiency for many systems, does not provide spectroscopic-quality vibrational frequencies. Moreover, studying complex bulk systems does not allow to disentangle all the subtle atomic interactions, which would be required for a thorough validation of the PES. 

In the present work, we fill this gap by investigating in detail the accuracy that can be achieved by HDNNPs as a typical example for second-generation MLPs. We explore the limits of this method, which has originally been designed for dealing with very large numbers of atoms, by carefully training a coupled-cluster-quality~\cite{bartlett2007} PES for a moderately sized system. This does not only allow us to compute harmonic frequencies but also to assess the quality of highly accurate anharmonic frequencies using state-of-the-art methods under well-controlled conditions. 

As a model system we have chosen the formic acid dimer (FAD), a doubly hydrogen-bonded complex, which in recent years has become a benchmark system for the development of molecular PESs and the calculation of accurate vibrational spectra thanks to the increasing body of gas-phase spectroscopic data\cite{HerGeoHeppHur,GEORGES2004187,ZieSu,XueSu,KoLaDoNoSu12,NeSu20,NeMeKoXuSu21}. It is the smallest hydrogen bonded complex with double proton transfer and as such it is a system of high interest for spectroscopy and theoretical dynamic studies due to the possible delocalization of the nuclei over the two wells. This double proton transfer is challenging since it cannot be described by standard near-equilibrium tools based on normal coordinates and perturbation theory. Even for this seemingly small ten-atom system and its 24-dimensional PES the determination of accurate anharmonic frequencies is computationally demanding and thus hardly accessible by a direct application of wave function electronic structure methods. For this reason the standard approach is the intermediate construction of analytic PES for the determination of vibrational frequencies.

Hence, in general, theoretical studies on vibrational frequencies have three limiting factors: the electronic structure method, its representation by a multi-dimensional PES function, and the vibrational treatment. Regarding the first two aspects, two ab initio-based analytic PESs have been proposed for FAD in the literature. In 2016, Qu and Bowman developed the first full-dimensional potential energy surface \cite{QuBo16} (henceforth labelled as QB16) for FAD by fitting permutationally invariant polynomials, a very accurate method providing PESs of a quality similar to modern MLPs~\cite{P2881}, to 13475 \emph{ab initio} energy points computed at the CCSD(T)-F12/haTZ level of electronic structure theory. Later, they carried out vibrational configuration interaction computations in normal coordinates \cite{QuBo18jcpl,QuBo18high,QuBo18fd} on this surface. Recently, two of us used the QB16 PES and tested, using a reduced dimensionality model, the utility of normal coordinates or a possible efficiency gain of using curvilinear (normal) coordinates \cite{DaAvMa21} in the GENIUSH program \cite{MaCzCs09}. During this work, several fundamental, combination, and overtone frequencies in the fingerprint range were obtained in an excellent agreement with experiment~\cite{NeSu20}. However, two (totally symmetric) fundamental frequencies were obtained strongly blueshifted in comparison with the harmonic frequencies of the QB16 PES\cite{QuBo16} and in comparison with the experimental value~\cite{NeSu20}. Based on these observations and due to some `artificial' features of the QB16 PES that made it necessary to restrict the quadrature grid used for the vibrational computations, it was concluded that some further work on improving the FAD PES is required.

In 2022, K\"aser and Meuwly reported another full-dimensional PES (PES$_{\mathrm{TL}}$) for FAD \cite{KaMe22} generated by the message passing neural network PhysNet~\cite{P5577}. It is based on 26,000 MP2/aug-cc-pVTZ single point energies, which have then been transfer-learned employing 866 CCSD(T)/aug-cc-pVTZ energies to obtain an approximately coupled cluster-quality PES that has also been used in the computation of harmonic and anharmonic vibrations. 
Regarding experiment, FAD is a very well-studied system (see, \eg\ Ref.~\citenum{NeSu20} and \citenum{NeMeKoXuSu21} and references therein), and a wealth of data is available for the validation of theoretical frequencies. 

In early work, thermal gas phase spectroscopy of FAD has been done~\cite{MiPi1958,BeMi}, while more recently  
jet-cooled infrared and Raman spectra of FAD in the monomer finger print region up to 1500 \cm\ have been studied in Ref.~\citenum{NeMeKoXuSu21}.
Over the past one and a half decades, all intermolecular vibrational fundamentals and several combination and overtone bands of FAD have been determined in the gas phase with an experimental uncertainty of 1~cm$^{-1}$\cite{GEORGES2004187,MatyRiehn,FumTai,BertieJohn,BiHa09,KoLaDoNoSu12,FumTai,OrtHa,ZieSu,XueSu,HerGeoHeppHur}. The experimental results including a critical evaluation of theoretical work still lagging behind experiment have been recently reviewed in Ref.~\citenum{NeSu20}.

The aim of the present work is to develop a robust and full-dimensional high-quality HDNNP for FAD and to benchmark the obtained frequencies using the best available theoretical and experimental data.
By robustness, we mean that the 24-dimensional (24D) hypersurface provides a faithful representation (possibly without `holes' or other unphysical features) of the \emph{ab initio} electronic energy points  over the relevant quantum dynamical range, and this robustness persists irrespective of the actual choice of (normal or curvilinear) internal coordinates.
By accuracy, in this context, we mean that the fitted hypersurface reproduces \emph{ab initio} points  sufficiently closely. `Sufficient' is determined in relation with prospective (ro)vibrational computations in comparison with (gas-phase) experimental infrared and Raman spectroscopy data.\cite{NeSu20}

Next to the RMSE, which is the most common quality measure of PESs, we make use of additional quantities in this work that allow further refinement of the potential energy surface for spectroscopic applications. For this purpose, first, we define an accuracy goal for the harmonic frequencies that are expected to be reproduced by the PES to within 10~\cm\ with respect to the \emph{ab initio} harmonic frequency values computed at the same level of electronic structure theory. Furthermore, to have a relatively compact assessment of the mode couplings representation in the PES, we test the second-order vibrational perturbation theory (VPT2) frequencies of the PES.

Finally, although a semi-rigid description, i.e., relatively small amplitude vibrations about an equilibrium structure of the PES, appears to be a good starting point for FAD,
the concerted proton tunneling of the double hydrogen bond qualifies this complex for the family of systems with multiple (two) large-amplitude motions. `High'-dimensional systems with multiple large-amplitude motions, \emph{i.e.,} motions in which nuclei are delocalized over multiple PES wells, are common in molecular systems and cannot be efficiently described by using perturbative methods developed about equilibrium structure properties (underlying the normal coordinate concept).
An efficient quantum dynamics description of these types of systems is currently an active and challenging field for methodological developments \cite{AvMa19,AvMa19b,AvPaCzMa20,WaCa20,FeLaScBeBa19,LiLiFeBa21,ChBeScNaLa22}. 
These developments can be tested and validated with respect to precise spectroscopy data, assuming that faithful and accurate PES representations for the molecular system is available.

After giving a brief summary of the employed methods in Section~\ref{sec:methods} and the computational details in Section~\ref{sec:comp_details}, the results are presented in Section~\ref{sec:results}. First, we assess the quality of the HDNNP, which is obtained by iteratively increasing the amount of reference data, until a converged potential is obtained. This PES is then characterized by its harmonic frequencies, which are compared to coupled cluster data. Finally, we report anharmonic frequencies obtained from VPT2 and reduced-dimensionality variational calculations, which allow for a direct validation of the MLP using accurate experimental data.

\section{Method}\label{sec:methods}

\subsection{High-Dimensional Neural Network Potentials}

High-dimensional neural network potentials (HDNNP) have been introduced by Behler and Parrinello in 2007~\cite{P1174} as the first type of MLP applicable to large condensed systems containing thousands of atoms. This is achieved by representing the potential energy $E$ of the system as a sum of atomic energies $E_i$, which depend on the local chemical environment up to a cutoff radius,
\begin{equation}
     E=\sum_{i=1}^{N_{\mathrm{atom}}}E_{i} \quad .
 \end{equation}
Each of these atomic energies is the output of an individual atomic feed-forward neural network describing the functional relation between the respective energy contribution and the local atomic structure. The weight parameters of these neural networks are determined in an iterative training process using known energies (and often also forces) obtained from reference electronic structure calculations. The weight parameters and architectures of all atomic neural networks of a given chemical element are constrained to be the same making the potential transferable to different system sizes.

To ensure the mandatory translational, rotational and permutational invariances of the HDNNP-PES, the local atomic environments are typically characterized by vectors of atom-centered symmetry functions (ACSF)~\cite{P2882} as geometric descriptors, which meet these requirements by construction and provide local structural fingerprints. The inclusion of a cutoff function ensures that the ACSFs smoothly decay to zero in value and slope at the cutoff radius, which is commonly chosen between 5 and 10 \AA{}.

Several extensions to second-generation HDNNPs have been introduced in recent years, like the consideration of long-range electrostatic interactions based on environment-dependent charges~\cite{P2962,P3132}, non-local charge transfer~\cite{P5932}, and magnetic interactions~\cite{P6057}. More details about HDNNPs and their properties can be found in several recent reviews~\cite{P4444,P5128,P6018}. In the present work a second-generation HDNNP relying on ``short-range'' atomic energies only is employed, since for a comparably small benchmark system like the FAD all interactions can be fully described as a function of the local chemical environments provided that a sufficiently large cutoff radius is chosen.

\subsection{Frequency calculations}

\subsubsection{Harmonic frequencies}

The determination of harmonic frequencies $\omega$ is a routine task in theoretical spectroscopic investigations.\cite{Wilson1980} Here, we construct the Cartesian second derivative matrix (Hessian) by means of finite differences based on the HDNNP. Mass weighting and diagonalization then yields the desired harmonic vibrational frequencies. These are then compared directly to the corresponding results of reference \abinit\ calculations.

\subsubsection{Second-order vibrational perturbation theory}

The anharmonic fundamental transition frequency $\nu_i$ within VPT2 is given by\cite{Clabo1988} 
\begin{equation}\label{eq:nui}
    \nu_i = \omega_i + 2x_{ii} + \frac{1}{2}\sum_{j{\neq}i} x_{ij}\,,
\end{equation}
where $\omega_i$ is the harmonic vibrational wavenumber and the $x_{ij}$ are the anharmonicity constants that account for anharmonicity in the vibrational mode $i$ as well as the coupling to other modes $j$. Equation~\ref{eq:nui} is based on a quartic force field (QFF), i.e.,\ a potential energy expression in terms of dimensionless normal coordinates $\mathbf{q}=\left\{q_i\right\}$ given by a Taylor expansion up to fourth order,\cite{Clabo1988}
\begin{equation}\label{eq:QFF}
    E(\mathbf{q}) = \frac{1}{2}\sum_i \omega_i q^2_i + \frac{1}{6}\sum_{ijk} \phi_{ijk} q_iq_jq_k + \frac{1}{24}\sum_{ijkl} \phi_{ijkl} q_iq_jq_kq_l\,.
\end{equation}
In Eq.~\ref{eq:QFF} the so-called cubic and quartic force constants are denoted by $\phi_{ijk}$ and $\phi_{ijkl}$, respectively. A contact transformation of the (ro)vibrational Hamiltonian\cite{Amat1971,Papousek1982,Aliev1985} up to second order then yields formulae for the $x_{ij}$ in terms of molecular parameters,
\begin{equation}\label{eq:xii}
  x_{ii} = \frac{1}{4}\phi_{iiii} - \frac{1}{16}\sum_{j}\phi^2_{iij}\frac{8\omega^2_i-3\omega^2_j}{\omega_j(4\omega^2_i-\omega^2_j)}
\end{equation}
and
\begin{align}\label{eq:xij}
  x_{ij} &= \frac{1}{4}\phi_{iijj} - \frac{1}{4}\sum_{k}{\phi_{iik}\phi_{kjj}}\frac{1}{\omega_k} - \frac{1}{2}\sum_{k}\phi_{ijk}\frac{\omega^2_k-\omega^2_i-\omega^2_j}{\Delta_{ijk}}\nonumber\\ 
  &+ \sum_\alpha B^\alpha_\mathrm{e}\zeta^\alpha_{ij}\zeta^\alpha_{ij}\left(\frac{\omega_i}{\omega_j}+\frac{\omega_j}{\omega_i}\right)\,,
\end{align}
where the resonance denominator $\Delta_{ijk}$ is given by
\begin{equation}\label{eq:dijk}
    \Delta_{ijk}=(\omega_i+\omega_j+\omega_k)(\omega_i+\omega_j-\omega_k)(\omega_i-\omega_j+\omega_k)(\omega_i-\omega_j-\omega_k)\,.
\end{equation}
The last term in Eq.~\eqref{eq:xij} depends on the three equilibrium rotational constants $B^\alpha_\mathrm{e}$ and the Coriolis coupling constants $\zeta^\alpha_{ij}$.

A closer look at Eq.~\eqref{eq:xii} and Eq.~\eqref{eq:xij} shows that they contain differences of harmonic frequencies in denominators which may lead to so-called Fermi-Resonances. Two cases need to be accounted for within VPT2: $\omega_i\approx2\omega_j$ and $\omega_i\approx\omega_j+\omega_k$. In fact, such vibrational resonances have been well documented experimentally for FAD (see Ref.~\citenum{NeMeKoXuSu21} and references therein) and therefore can be expected to interfere in the present VPT2 treatment. In order to allow a comparison between VPT2 results and experiment a special treatment is required which is sometimes referred to as GVPT2.\cite{GVPT2} First, a change in the contact transformation removes the resonance denominators from the $x_{ij}$ to yield so-called \textit{deperturbed} anharmonicity constants $x^*_{ij}$.\cite{Nielsen1959} In a second step the resonating vibrational states are treated using perturbation theory for (near) degenerate states\cite{Califano1976} where the coupling matrix element depends on a cubic force constant, \ie\ $\phi_{ijj}$ or $\phi_{ijk}$.

\subsubsection{Reduced-dimensionality variational vibrational computations \label{sec:var}}

A complete quantum dynamical characterization of a molecular system can be obtained by the variational solution of the (ro)vibrational Schrödinger equation including the multi-dimensional PES as an `effective' interaction acting among the nuclei.
The vibrational Hamiltonian, as a sum of the PES, $E$, 

and the kinetic energy operator written in general coordinates is 
\begin{align}
  \hat{H} 
  = 
  \frac{1}{2} \sum_{k=1}^D \sum_{l=1}^D \tilde{{g}}^{-1/4} \hat{p}_k G_{kl}\tilde{{g}}^{1/2}  \hat{p}_l\tilde{{g}}^{-1/4}  + E
\end{align}
where
$\hat{p}_k=-i\partial/ \partial q_k$ ($k=1,\ldots,D\leq 3N_{\mathrm{atom}}-6$), 
and 
the $q_k$ vibrational coordinate definition (and choice of the body-fixed frame) is encoded in the mass-weighted metric tensor,
$\pmb{g}\in\mathbb{R}^{D\times D}$, with $\tilde g=\text{det}\pmb{g}$ and $\pmb{G}=\pmb{g}^{-1}$. 

Rigorous geometrical constraints ($D<3N_{\mathrm{atom}}-6$) of selected internal coordinates can be introduced in the quantum mechanical vibrational model by `deleting' the relevant rows and columns of the $\pmb{g}$ matrix. 

A numerical kinetic energy operator (KEO) approach based on this formalism has been implemented in the GENIUSH computer program\cite{MaCzCs09}, \emph{i.e.,} $\pmb{g}$, $\tilde{g},$ and $\pmb{G}$ are evaluated at grid points. The original implementation relied on the discrete variable representation (DVR) \cite{LiCa00} of the vibrational Hamiltonian, but more recently, this numerical KEO approach has been used with finite basis representation (FBR) and the Smolyak scheme \cite{AvMa19,AvMa19b,DaAvMa22}, which opens the route towards higher-dimensional computations.
For FAD, the lowest-energy vibrational frequencies from the fingerprint range have been converged for a series of reduced-dimensionality vibrational models defined in Ref.~\citenum{DaAvMa21}.

\begin{figure}!
    \centering
    \includegraphics[scale=0.39]{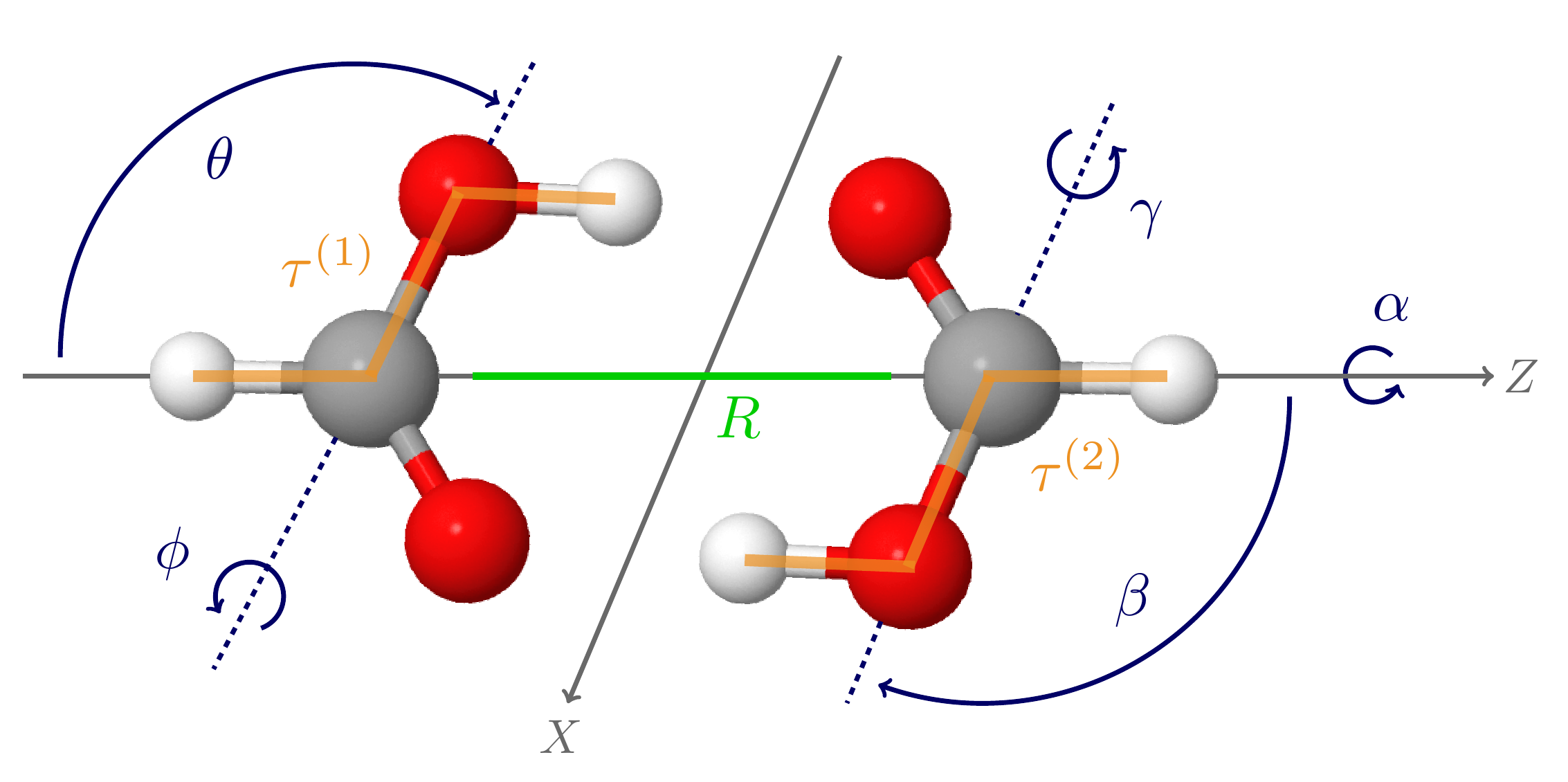}
    \caption{Formic acid dimer in its equilibrium structure. The intermolecular ($R,\theta,\phi,\alpha,\beta,\gamma$) and the two cis-trans torsional ($\tau^{(1)}$ and $\tau^{(2)}$) coordinates used in the 8-dimensional variational vibrational computations are also shown.}
    \label{fig:fadcoor}
\end{figure}

The first applications with the new HDNNP-PES developed in the present work
use an 8-dimensional (8D) vibrational model, including the six intermolecular modes and the cis-trans torsional degrees of freedom of both monomers (Figure~\ref{fig:fadcoor}), which was found to perform reasonably well in Ref.~\citenum{DaAvMa21}.

\section{Computational Details}\label{sec:comp_details}

The reference coupled cluster calculations for generating the training data set have been performed using \textsc{Molpro} 2019\cite{MOLPRO}. The explicitly correlated frozen-core CCSD(T)-F12a\cite{CCa1,CCa2,CCa3} method has been used in conjunction with an aug-cc-pVTZ basis\cite{avnz} for carbon and oxygen atoms and a cc-pVTZ basis\cite{vnz} for hydrogen atoms. In the following this atomic orbital basis will be abbreviated haTZ. 

A VTZ/JKFIT basis\cite{jkfit} has been employed for the resolution of identity approximation. Reference \abinit{} equilibrium geometries and harmonic frequencies were

obtained using either \textit{default} convergence criteria or a \textit{tight} definition which corresponds to: a lower threshold for screening of two-electron integrals ($\textsc{twoint}=10^{-16}$) and 
the energy ($\textsc{energy}=10^{-10}$) for all CCSD(T)-F12a energy evaluations, improved convergence of the geometry optimization ($\textsc{gradient}=10^{-6}$ and $\textsc{step}=10^{-6}$) employing a \textsc{fourpoint} numerical gradient and finally reduced step sizes of 0.005~$a_{0}$ for both optimization and numerical Hessian.
The HDNNPs have been constructed using the program RuNNer~\cite{P5128,P4444}. Several architectures of the atomic neural networks have been tested containing two hidden layers with up to 14 neurons each. A large cutoff radius of 15.0 Bohr has been used to ensure that in case of the FAD all atoms are included in all atomic environments. The parameters of the atom-centered symmetry functions describing the atomic environments are given in the Supporting Information. For the training process, the available data has been randomly split into a training set (90~\%) and an independent test set (10~\%) not used in the iterative weight optimization, which employs a global adaptive extended Kalman filter~\cite{P1308}.

The reference data set has been generated in several steps. An initial set of HDNNPs has been constructed using the energies of 13,475 structures from the work of Qu and Bowman\cite{QuBo18high}. This data has then been further extended by including additional structures obtained from active learning, i.e., by comparing the energy predictions of different HDNNPs. If for a given structure the deviation between these predictions was above a specified error threshold, a CCSD(T) calculation has been carried out for the respective structure, which has then been added to the training set to further refine the potential. 
Several strategies have been employed to search for structures not well-represented, which are geometries displaced along the 24 normal modes, geometries used in the numerical calculation of the Hessian, and structures obtained from molecular dynamics (MD) simulations at 100 and 300~K driven by preliminary intermediate HDNNPs using the n2p2~\cite{n2p2} and LAMMPS~\cite{P4473} codes. Moreover, about 500,000 structures corresponding to the direct product grid used in the variational vibrational computations\cite{DaAvMa21} were screened systematically and added if needed. A threshold for the predicted energy deviation of 1~meV/atom has been applied in the active learning for the MD simulations, while 0.02 eV/atom have been used for selecting geometries from the pool of grid structures. In total, this extended second data set contains the energies of 27,372 FAD structures. 
Finally, a third data set has been constructed by adding another 1,800 structures extracted from two-dimensional cuts of the PES corresponding to pairwise coupled harmonic modes, for which a threshold of 2 meV/atom has been used. 

\section{Results}\label{sec:results}

\subsection{High-Dimensional Neural Network Potentials}

\begin{table}[h!]
    \centering
        \caption{Energy root mean squared errors (RMSE) of the training and test sets for the three HDNNPs trained using data sets containing increasing numbers of structures. The RMSEs of these potentials are given for the complete energy range covered in reference data and for the structures in the most relevant energy range below 0.1 Ha with respect to the global minimum. The numbers of structures included in the respective energy range for calculating the RMSEs is given in the second column while in both cases the HDNNPs have been trained to the full data range.\label{tab:rmse}}
    \begin{tabular}{cccccc} 
    \toprule
         &   & \multicolumn{2}{c}{RMSE [meV/atom]}& \multicolumn{2}{c}{RMSE [\cm]} \\
         \cmidrule(lr){3-4} \cmidrule(lr){5-6}
    PES       & structures & training & testing & training & testing\\
    \midrule
    \multicolumn{6}{c}{full energy range}\\
            \cmidrule(lr){1-6}
      HDNNP1   &  13475 &  2.43        &    13.99     &    196 & 1129 \\
      HDNNP2   &  27372 &  0.92         &      1.88    &   74  &   158  \\
      HDNNP3   &  29162 &  0.37        &       2.04   &   30  &   165  \\
    \midrule
          \multicolumn{6}{c}{energy range below 0.1 Ha}\\
                      \cmidrule(lr){1-6}
      HDNNP1   &  12725 &  2.15    &  2.42 &  174 & 195  \\
      HDNNP2   &  26531 &  0.85    & 1.04  &  68  & 83   \\
      HDNNP3   &  28286 &  0.35    & 0.34  &  28  &  27  \\ 
    \bottomrule
    \end{tabular}
\end{table}

\begin{figure}[!]
\includegraphics[width=\linewidth]{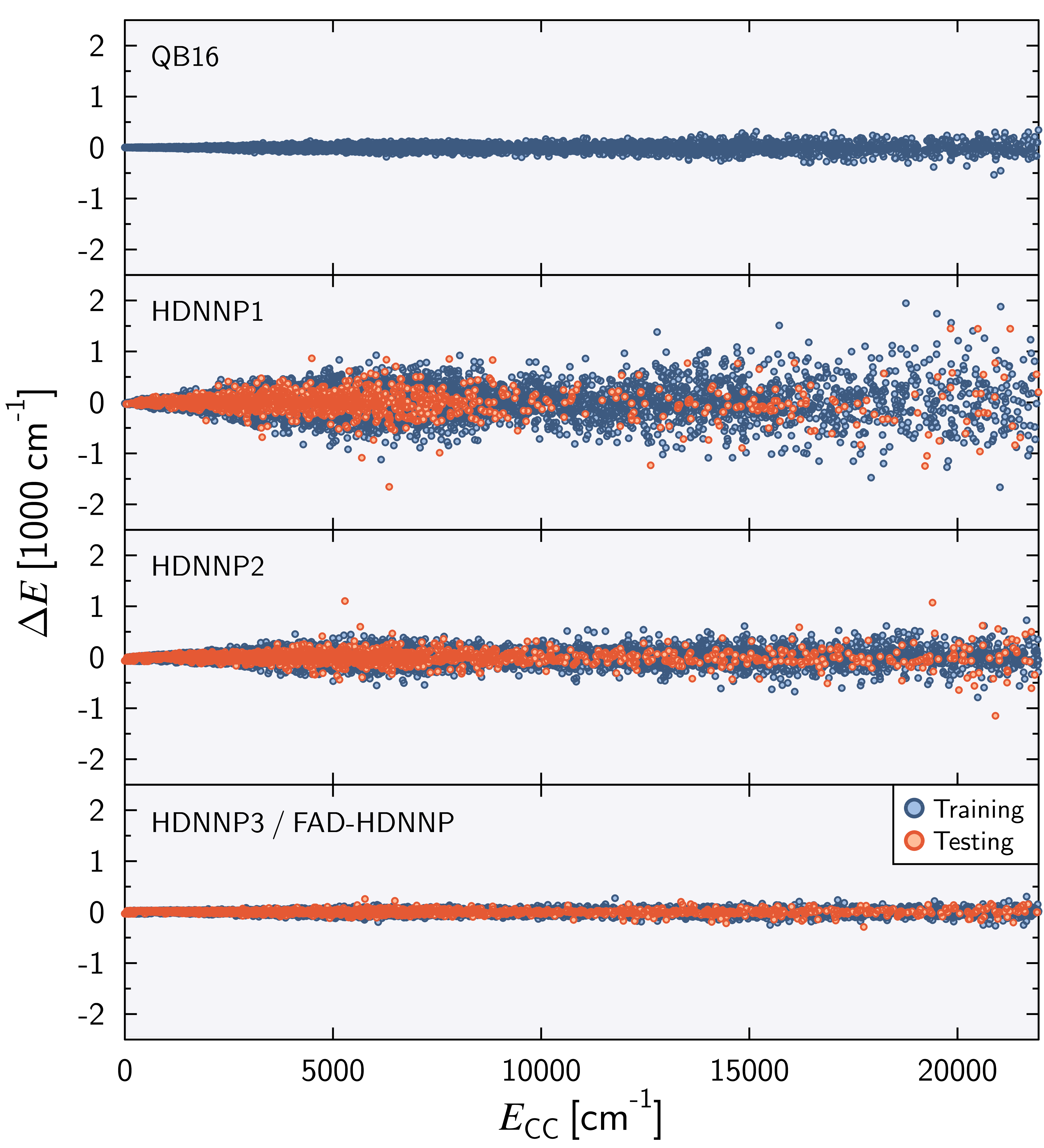}
\caption{Energy difference $\Delta E=E_\text{CC}-E_{\text{PES}}$ as a function of the reference energy $E_{\text{CC}}$ relative to the energy of the global minimum for the QB16 PES\cite{QuBo16}, HDNNP1, HDNNP2, and HDNNP3 (=FAD-HDNNP). The root-mean-squared errors (RMSE) for the HDNNPs are provided in Tab.~\ref{tab:rmse}.
}
\label{fig:fitting}
\end{figure}

The HDNNPs have been trained using the three different data sets containing increasing numbers of structures as described in the previous Section. The RMSEs of the energies of the resulting HDNNPs showing the best performance called HDNNP1 (11 neurons per hidden layer), HDNNP2 (10 neurons per hidden layer), and HDNNP3 (14 neurons per hidden layer) are compiled in Table~\ref{tab:rmse}. Since most of the energies of the Qu and Bowman data set are within 0.1 Ha with respect to the energy of the global minimum geometry, the high-energy region is only sparsely sampled. This is the reason for the large test set error of all HDNNPs compared to the respective error of the training set, indicating overfitting in the high-energy region beyond 0.1 Ha, which is particularly pronounced for HDNNP1 relying on the data of Qu and Bowman only. This overfitting is not present in the very well sampled low-energy region below 0.1 Ha, as can be seen in the bottom half of Table~\ref{tab:rmse}.

Figure~\ref{fig:fitting} shows the energy error of all training and test data points for the QB16 potential and the three HDNNPs. The QB16 potential performs better than \pesdevel{1} if only the original data of Qu and Bowman is used. The (unweighted) RMSE of the QB16 on this initial data set corresponds to 0.91~meV/atom (74~cm$^{-1}$) to be compared with 2.43~meV/atom (196~cm$^{-1}$) for \pesdevel{1}. If, however, the data set is increased, the error of the HDNNPs is strongly reduced finally resulting in a very small RMSE of only about 0.35~meV/atom (28~cm$^{-1}$) for HDNNP3 in the relevant energy range up to 0.1~Ha (21950~cm$^{-1}$).

\subsection{Vibrational frequencies}

\subsubsection{Harmonic frequencies}

\begin{figure}[!]
\includegraphics[width=\linewidth]{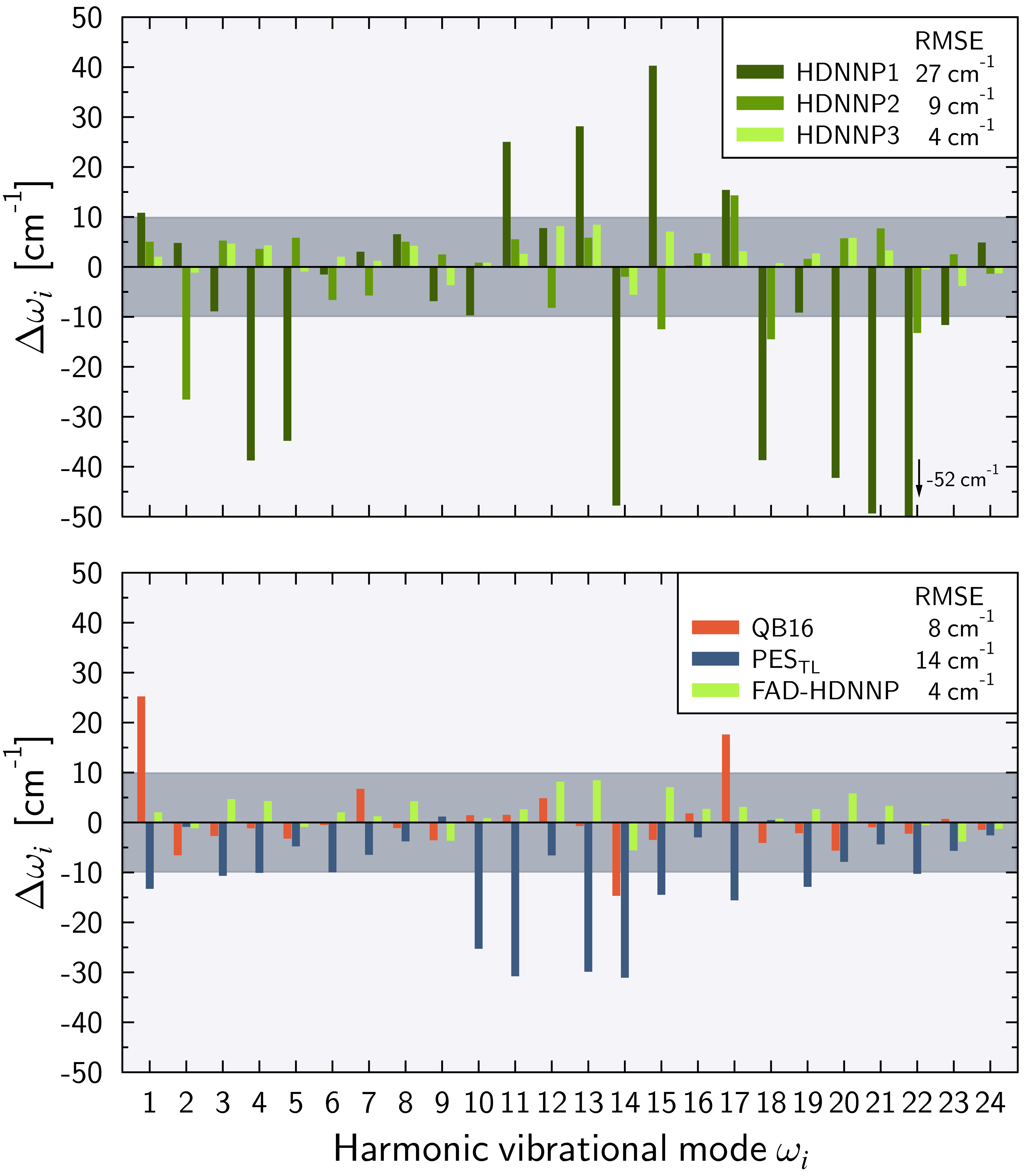}
\caption{Deviations of the harmonic vibrational frequencies $\omega_i$ with respect to the reference \abinit\ results (\textit{tight} settings). Top panel depicts the deviations $\Delta \omega_i = \omega_i - \omega_{\text{ref},i}$ for different HDNNP versions during the potential refinement with HDNNP3 corresponding to the final FAD-HDNNP. The bottom panel compares these differences for harmonic frequencies calculated from the QB16 PES by Qu and Bowman\cite{QuBo16}, the harmonic frequencies published by K\"aser and Meuwly\cite{KaMe22} based on the transfer-learned PES\textsubscript{TL}, and the final FAD-HDNNP (= HDNNP3) results. The reference \abinit\ results have been obtained at the level of electronic structure theory that was used for the development of the respective PES, \ie\ fc-CCSD(T)/AVTZ for PES\textsubscript{TL} (see Ref.~\citenum{KaMe22} for details) and fc-CCSD(T)-F12a/haTZ for QB16 as well as FAD-HDNNP. Root mean squared errors (RMSE) are provided in the legends.}
\label{fig:hotest}
\end{figure}

\begin{table}[!]
    \centering
    \caption{Comparison of the harmonic frequencies $\omega_i$ (in cm$^{-1}$) for the three HDNNPs and the respective reference \abinit\ results obtained at the fc-CCSD(T)/haTZ level of theory. The latter are computed using either the standard convergence criteria (termed `default') or stricter convergence criteria (termed `tight') with respect to the electronic energy and geometry optimization as well as smaller step sizes in the numerical gradient and Hessian evaluation (cf.~Sec.~\ref{sec:comp_details})}
    \label{tab:frq_nnps_table}
    \begin{tabular}{llp{1.3cm}p{1.3cm}p{1.3cm}p{1.0cm}p{1.0cm}} 
    \toprule
               &       &        &       & \pesfinal & \multicolumn{2}{c}{CCSD(T)-F12a}\\
                                                  \cmidrule(lr){6-7}
          Mode &  Sym. &   HDNNP1 &  HDNNP2 &  HDNNP3 &  default & tight\\
    \midrule
    $\omega_1$ & A$_g$ &   3218 &  3212 &  3209 &  3203 &  3207 \\
    $\omega_2$ & A$_g$ &   3060 &  3077 &  3103 &  3105 &  3103 \\
    $\omega_3$ & A$_g$ &   1708 &  1723 &  1722 &  1717 &  1718 \\
    $\omega_4$ & A$_g$ &   1443 &  1485 &  1486 &  1484 &  1482 \\
    $\omega_5$ & A$_g$ &   1376 &  1417 &  1410 &  1413 &  1411 \\
    $\omega_6$ & A$_g$ &   1254 &  1249 &  1257 &  1257 &  1255 \\
    $\omega_7$ & A$_g$ &    689 &   681 &   688 &   688 &   686 \\
    $\omega_8$ & A$_g$ &    216 &   215 &   214 &   211 &   210 \\
    $\omega_9$ & A$_g$ &    164 &   173 &   167 &   171 &   170 \\[1ex]
 $\omega_{10}$ & B$_g$ &   1073 &  1083 &  1083 &  1085 &  1083 \\
 $\omega_{11}$ & B$_g$ &    980 &   960 &   957 &   960 &   955 \\
 $\omega_{12}$ & B$_g$ &    257 &   241 &   257 &   258 &   249 \\[1ex]
 $\omega_{13}$ & A$_u$ &   1128 &  1106 &  1109 &  1102 &  1101 \\
 $\omega_{14}$ & A$_u$ &    937 &   983 &   979 &   986 &   985 \\
 $\omega_{15}$ & A$_u$ &    213 &   161 &   180 &   186 &   173 \\
 $\omega_{16}$ & A$_u$ &     68 &    71 &    71 &    76 &    68 \\[1ex]
 $\omega_{17}$ & B$_u$ &   3324 &  3323 &  3312 &  3305 &  3309 \\
 $\omega_{18}$ & B$_u$ &   3108 &  3085 &  3100 &  3101 &  3099 \\
 $\omega_{19}$ & B$_u$ &   1773 &  1783 &  1784 &  1782 &  1782 \\
 $\omega_{20}$ & B$_u$ &   1411 &  1459 &  1459 &  1456 &  1453 \\
 $\omega_{21}$ & B$_u$ &   1358 &  1414 &  1410 &  1405 &  1407 \\
 $\omega_{22}$ & B$_u$ &   1208 &  1247 &  1260 &  1260 &  1260 \\
 $\omega_{23}$ & B$_u$ &    704 &   718 &   712 &   716 &   715 \\
 $\omega_{24}$ & B$_u$ &    281 &   275 &   275 &   278 &   276 \\ 
    \bottomrule
    \end{tabular}
\end{table}

\begin{table*}[!]
    \caption{Geometrical parameters of the formic acid dimer minimum structure optimized at the reference 
    \abinit\ level of theory and determined for different PESs. Bond lengths are provided in {\AA}ngstr\"oms and angles in degrees. \pesdevel{3} corresponds to the final \pesfinal\ for spectroscopic use.
    }
    \label{tab:geom}
    \centering
    \begin{tabular}{l*{6}{D{.}{.}{4}}} 
    \toprule
    &\multicolumn{2}{c}{\textit{Ab initio}}\\
    \cmidrule(lr){2-3}
    Parameter & 
    \multicolumn{1}{r}{default} & \multicolumn{1}{r}{tight} & 
    \multicolumn{1}{r}{QB16\com{a}} & \multicolumn{1}{r}{\pesdevel{1}} & \multicolumn{1}{r}{\pesdevel{2}} & \multicolumn{1}{r}{\pesdevel{3}} \\
    \midrule
    $r$(O--H)               &  0.9934 &  0.9932 &  0.9927 &  0.9945 &  0.9925 &  0.9936\\
    $r$(C--H)               &  1.0929 &  1.0930 &  1.0929 &  1.0927 &  1.0937 &  1.0927\\
    $r$(C--O)               &  1.3113 &  1.3114 &  1.3116 &  1.3104 &  1.3121 &  1.3112\\
    $r$(O$\cdots$O)         &  2.6748 &  2.6758 &  2.6778 &  2.6729 &  2.6791 &  2.6709\\
    $r$(C=O)                &  1.2177 &  1.2176 &  1.2174 &  1.2192 &  1.2172 &  1.2178\\
    $\angle$O=C--O          &126.14   &126.14   &126.15   &126.26   &126.13   &126.13\\
    $\angle$O=C--H          &122.02   &122.03   &122.05   &122.04   &122.15   &121.99\\
    $\angle$C--O--H         &109.77   &109.76   &109.73   &109.42   &109.93   &109.79\\
    $\angle$O--H${\cdots}$O &178.93   &178.93   &178.95   &179.51   &179.01   &178.86\\
    \bottomrule
    \addlinespace[0.25ex]

    \multicolumn{6}{l}{\com{a} Qu and Bowman\cite{QuBo16}}
    \end{tabular}
\end{table*}

As an important test for spectroscopic applications, next we compare the harmonic frequencies of the PESs with the respective \abinit\ values corresponding to the same level of electronic structure theory. As described in the introduction we consider an agreement within $\pm 10$~\cm\ as a requirement for this purpose.
Figure~\ref{fig:hotest} shows the harmonic frequency errors obtained for the three main HDNNPs of the PES refinement procedure. Similar to the energy RMSEs we find a continuous improvement starting from HDNNP1 showing some rather large deviations up to 52~\cm, which are decreasing to at most 27~\cm\ for HDNNP2, finally reaching a very high quality in case of HDNNP3 with with the largest deviation in the PES vs. \emph{ab initio} harmonic frequencies being less than 7~\cm. Overall, the frequency RMSEs of HDNNP1, HDNNP2, and HDNNP3 are about 27~\cm, 9~\cm, and 4~\cm, respectively. Consequently, we choose HDNNP3 as the production-quality PES for vibrational calculations from now on called simply `FAD-HDNNP'.

The lower part of Fig.~\ref{fig:hotest} provides a comparison of the FAD-HDNNP with the other two published FAD PESs.\cite{QuBo16,KaMe22} Concerning the accuracy of the PES at the harmonic level, \pesfinal\ seems to outperform the earlier QB16\cite{QuBo16} and Phys$_\text{TL}$\cite{KaMe22} potentials, which exhibit RMSEs of 8~\cm\ and 14~\cm\, with maximum deviations of 28 and 30~cm$^{-1}$, respectively, from the \abinit\ harmonic frequency values of the corresponding level of electronic structure theory.

The numerical values of the harmonic frequencies for the newly developed \pesfinal\ (and also for the \pesdevel{1}\ and \pesdevel{2}\ development stages) are collected in Table~\ref{tab:frq_nnps_table} together with the \abinit\ frequencies. Interestingly, the numerical uncertainty of the \abinit\ values at the chosen level of theory still depends notably on the precise details of the \textsc{Molpro} computation setup as shown in the last two columns of the table. The current 4~cm$^{-1}$ RMSE with a maximum deviation of 7~cm$^{-1}$ of \pesfinal\ appears to have reached the current accuracy limit of the available electronic structure methodology employing commonly used default settings, as even the \textsc{Molpro} frequencies exhibit changes up to 13~\cm in exceptional cases like $\omega_{15}$ when using tight settings. Still, even in this case the RMSE of FAD-HDNNP with respect to the tight \textsc{Molpro} data is very similar to the RMSE with respect to the default settings with only a marginally increased maximum deviation of 8~\cm in case of $\omega_{12}$ and $\omega_{13}$.

Finally, the equilibrium structure of FAD obtained from the reference \abinit{} calculations can be compared to the minima of the HDNNPs. This is done in Table~\ref{tab:geom} and additional results obtained with the QB16 PES\cite{QuBo16} are provided for comparison. Due to symmetry only 9 internal coordinates are required to characterize FAD. Similar to the harmonic frequencies, the agreement of the HDNNP geometrical parameters with the reference \abinit{} results improves with the the refinement procedure. The largest difference is observed for the $r$(O$\cdots$O) distance which at the same time appears to be the most sensitive coordinate exemplified by a comparably large variation of 0.001~\AA{} between the default and tight reference \abinit{} geometries. Nevertheless, the agreement of \pesdevel{3} (=\pesfinal{}) with the (tight) reference \abinit{} geometry is good with an RMSE of 0.002~\AA\ for the bond distances and 0.04$^\circ$ for the angular coordinates.

\subsubsection{VPT2 frequencies\label{sec:vpt2num}}
\begin{table}[!]
    \centering
    \caption{%
      Comparison of the VPT2 fundamental frequencies (in cm$^{-1}$) obtained from the FAD-HDNNP with experimental data~\cite{BeMi,GEORGES2004187,ItoNaka,XueSu,NeMeKoXuSu21}.
    }
    \label{tab:frq_vpt2}
    \begin{tabular}{llp{0.3cm}rp{0.3cm}r}
\toprule
       Mode &  Sym. && FAD-HDNNP && Exp.\\
\midrule
    $\nu_1$ & A$_g$ &&    2920$\hphantom{\text{\com{a}}}$  && 2949$\hphantom{\text{\com{a}}}$\\
    $\nu_2$ & A$_g$ &&    2948$\hphantom{\text{\com{a}}}$  && 2900$\hphantom{\text{\com{a}}}$\\
    $\nu_3$ & A$_g$ &&    1677\com{a} && 1664\com{b}\\
    $\nu_4$ & A$_g$ &&    1433$\hphantom{\text{\com{a}}}$  && 1430$\hphantom{\text{\com{a}}}$\\
    $\nu_5$ & A$_g$ &&    1375$\hphantom{\text{\com{a}}}$  && 1375$\hphantom{\text{\com{a}}}$\\
    $\nu_6$ & A$_g$ &&    1229$\hphantom{\text{\com{a}}}$  && 1224$\hphantom{\text{\com{a}}}$\\
    $\nu_7$ & A$_g$ &&     682$\hphantom{\text{\com{a}}}$  &&  681$\hphantom{\text{\com{a}}}$\\
    $\nu_8$ & A$_g$ &&     197$\hphantom{\text{\com{a}}}$  &&  194$\hphantom{\text{\com{a}}}$\\
    $\nu_9$ & A$_g$ &&     164$\hphantom{\text{\com{a}}}$  &&  161$\hphantom{\text{\com{a}}}$\\[1ex]
 $\nu_{10}$ & B$_g$ &&    1058$\hphantom{\text{\com{a}}}$  && 1058$\hphantom{\text{\com{a}}}$\\
 $\nu_{11}$ & B$_g$ &&     934$\hphantom{\text{\com{a}}}$  &&  911$\hphantom{\text{\com{a}}}$\\
 $\nu_{12}$ & B$_g$ &&     247$\hphantom{\text{\com{a}}}$  &&  242$\hphantom{\text{\com{a}}}$\\[1ex]
 $\nu_{13}$ & A$_u$ &&    1074$\hphantom{\text{\com{a}}}$  && 1069$\hphantom{\text{\com{a}}}$\\
 $\nu_{14}$ & A$_u$ &&     964\com{c} &&  939\com{d}\\
 $\nu_{15}$ & A$_u$ &&     166$\hphantom{\text{\com{a}}}$  &&  168$\hphantom{\text{\com{a}}}$\\
 $\nu_{16}$ & A$_u$ &&      68$\hphantom{\text{\com{a}}}$  &&   69$\hphantom{\text{\com{a}}}$\\[1ex]
 $\nu_{17}$ & B$_u$ &&    3041$\hphantom{\text{\com{a}}}$  && 3050$\hphantom{\text{\com{a}}}$\\
 $\nu_{18}$ & B$_u$ &&    2941$\hphantom{\text{\com{a}}}$  && 2940$\hphantom{\text{\com{a}}}$\\
 $\nu_{19}$ & B$_u$ &&    1745$\hphantom{\text{\com{a}}}$  && 1741$\hphantom{\text{\com{a}}}$\\
 $\nu_{20}$ & B$_u$ &&    1416$\hphantom{\text{\com{a}}}$  && 1407$\hphantom{\text{\com{a}}}$\\
 $\nu_{21}$ & B$_u$ &&    1375$\hphantom{\text{\com{a}}}$  && 1372$\hphantom{\text{\com{a}}}$\\
 $\nu_{22}$ & B$_u$ &&    1233\com{e} && 1234\com{f}\\
 $\nu_{23}$ & B$_u$ &&     706$\hphantom{\text{\com{a}}}$  &&  708$\hphantom{\text{\com{a}}}$\\
 $\nu_{24}$ & B$_u$ &&     264$\hphantom{\text{\com{a}}}$  &&  264$\hphantom{\text{\com{a}}}$\\
\bottomrule
    \addlinespace[0.5ex]
    \multicolumn{6}{p{7.6cm}}{\com{a} Fermi-resonance coupled with $\nu_{4}+\nu_{8}$ at 1625~cm$^{-1}$.}\\
    \multicolumn{6}{p{7.6cm}}{\com{b} Experimental bands at 1664 and 1668~cm$^{-1}$}\\
    \multicolumn{6}{p{7.6cm}}{\com{c} Fermi-resonance coupled with $\nu_{12}+\nu_{23}$ at 941~cm$^{-1}$.}\\
    
    \multicolumn{6}{p{7.6cm}}{\com{d} Experimental bands at 939 and 953~cm$^{-1}$.}\\
    \multicolumn{6}{p{7.6cm}}{\com{e} Fermi-resonance coupled with $\nu_{10}+\nu_{15}$ at 1219~cm$^{-1}$.}\\
    \multicolumn{6}{p{7.6cm}}{\com{f} Experimental bands at 1220, 1225, and 1234~cm$^{-1}$.}
    
    \end{tabular}
\end{table}
Since the harmonic frequencies are well described by the \pesfinal, it is now appropriate to investigate the representation of mode couplings as a next step. A compact measure for the mode couplings near the equilibrium structure can be obtained from a comparison of the VPT2 frequencies computed using the PES with gas-phase experimental results\cite{BeMi,GEORGES2004187,ItoNaka,XueSu,NeMeKoXuSu21}. The corresponding anharmonic frequencies are provided in Table~\ref{tab:frq_vpt2}. The normal coordinate QFF parameters obtained from \pesfinal\ are compiled in the Supplementary Information, and can be used for further assessment of the representation of the mode couplings. This allows a comparison which is completely unaffected by details of the VPT2 resonance treatment.

Overall the agreement between \pesfinal\ and the experimental results is reasonable with an RMSE of 14~cm$^{-1}$, which is in the typical range of errors that are to be expected for the underlying level of \abinit{} theory. To improve on this would require the inclusion of high-level corrections such as core-valence correlation and higher-order correlation beyond CCSD(T). While such composite schemes have been shown to provide high-quality potentials for small molecules\cite{CVRQD,FPD2012,GaWeFoLe2021,Schr2022}, their computational cost is prohibitive for FAD. 

When comparing the VPT2-based results for the high-frequency hydrogen-stretching vibrations $\nu_1$, $\nu_2$, $\nu_{17}$, and $\nu_{18}$ with experiment relatively large differences are found. This is due to the fact that these modes are in an energy range of high state density which leads to substantial anharmonic couplings/resonances beyond what can be reliably treated using VPT2. Considering this, the RMSE without accounting for these high-frequency fundamentals approximately halves to 9~cm$^{-1}$. Nevertheless, some results warrant a more detailed discussion, especially those where resonance effects are present.

For $\nu_{14}$ we observe a large deviation of 25~cm$^{-1}$ when comparing the \pesfinal\ and the experimental frequency of 964 and 939~cm$^{-1}$, respectively. In agreement with experimental results\cite{MeSu,NeMeKoXuSu21} we find this mode to be in Fermi-resonance with $\nu_{12}+\nu_{23}$ at 941~cm$^{-1}$ to be compared to the experimental value of 961~cm$^{-1}$. A close look at these numbers may indicate a possible misassignment. However, upon inspection of the VPT2 results we find that these vibrational states are in an almost perfect resonance (56:44 mixing) with deperturbed energies of 952 and 954~cm$^{-1}$ for $(\nu_{12}+\nu_{23})^*$ and $\nu^*_{14}$, respectively. As such this resonance is very sensitive to the level of theory and small changes in the PES can easily reverse the state ordering.
Finally, our calculations indicate a Fermi-resonance between $\nu_{22}$ and $\nu_{10}+\nu_{15}$ with anharmonic transition frequencies of 1234 and 1219~cm$^{-1}$, respectively. Nejad~\textit{et al.} have discussed this resonance in detail,\cite{NeMeKoXuSu21} proposing a new assignment for the transitions in the triplet of bands centered at 1220, 1225, and 1234~cm$^{-1}$, i.e.\ a resonance triad assigned to $\nu_{10}+\nu_{15}$, $\nu_{9}+\nu_{11}+\nu_{15}$ and $\nu_{22}$. In contrast, we find no involvement of $\nu_{9}+\nu_{11}+\nu_{15}$ in the VPT2 results for $\nu_{22}$. Furthermore, the previously proposed resonance between $\nu_{10}+\nu_{15}$ and $\nu_{9}+\nu_{11}+\nu_{15}$ would require also for the $\nu_{10}$ fundamental to be in resonance with $\nu_{9}+\nu_{11}$, which we again do not observe in our VPT2 calculations. Moreover, our calculations yield virtually the same $\nu_{10}$ transition frequency compared to experiment. Therefore, the agreement of the VPT2 based $\nu_{10}+\nu_{15}$ and $\nu_{22}$ frequencies (1219 and 1233~cm$^{-1}$) with two of the three experimentally observed bands (1220 and 1235~cm$^{-1}$) appears to rule out the proposed assignment of $\nu_{9}+\nu_{11}+\nu_{15}$.

The latter band is predicted by VPT2 applied to \pesfinal{} at 1255~cm$^{-1}$. Based on our calculations the third observed band at 1225~cm$^{-1}$ may be assigned to $\nu_{23}+2\nu_{24}$ for which we obtain 1219~cm$^{-1}$. This assignment has also been previously proposed by Goroya and coworkers.\cite{GoZhSuDu}

\subsubsection{Reduced-dimensionality variational vibrational frequencies}

\begin{figure}[!]
    \centering
    \includegraphics[scale=0.55]{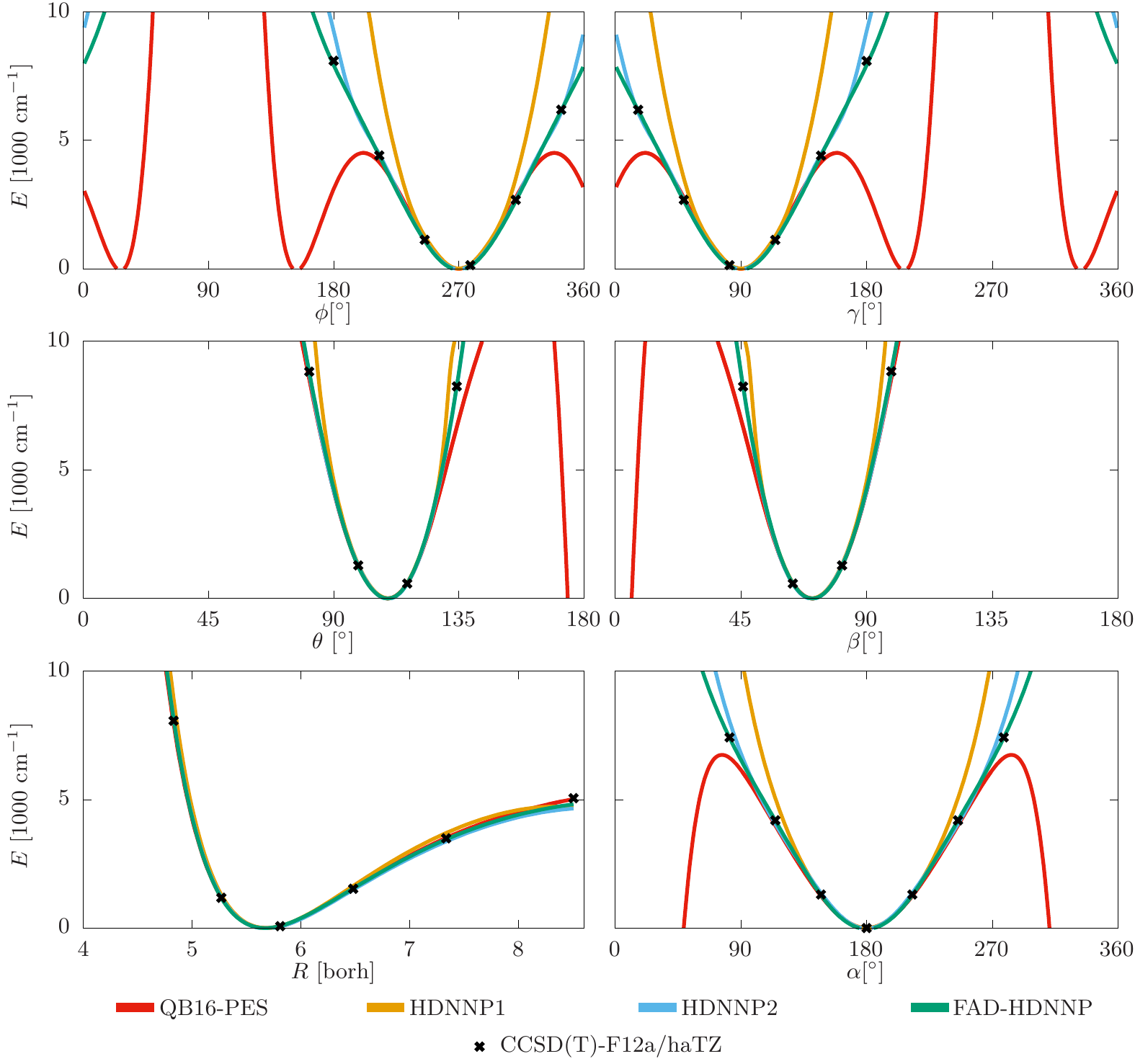}
    \caption{1-dimensional cuts of the three investigated HDNNPs and the QB16 potential along the intermolecular coordinates (cf. Fig.~\ref{fig:fadcoor}) of the formic acid dimer used in the reduced-dimensionality variational computations.
    }
    \label{fig:1dpescut}
\end{figure}

\begin{table}[!]
  \caption{%
    Vibrational energies referenced to the zero-point vibrational energy (in \cm)
    obtained with the 8D($\mathcal{I}$t) intermolecular-torsional model in the GENIUSH program using the QB16 PES and FAD-HDNNP.
     The inactive degrees of freedom are fixed at their respective equilibrium values. 
    \label{tab:8Dvib}  }
    \centering
    \begin{tabular}{@{}cccccc@{}}
\toprule
\multicolumn{1}{l}{Assignment\cite{DaAvMa21}} &
\multicolumn{1}{c}{$\tilde\nu_\text{QB16}$ \cite{DaAvMa21}}	&	
\multicolumn{1}{c}{$\tilde\nu_\text{FAD-HDNNP}$}	&	
\multicolumn{1}{c}{$\tilde\nu_\text{expt}$ \cite{NeSu20}}\\
\midrule
$\nu_{16}$          &	70	& 70	& 69.2	\\
$2\nu_{16}$         &	141	& 140	& 139	\\
$\nu_{15}$          &	162	& 171	& 168.5	\\
$\nu_{9}/\nu_{8}$   &	191	&	190	& 161 \\
$\nu_{8}/\nu_{9}$   &	208	&	210	& 194 \\
$3\nu_{16}$         &	211	&	210	&  	    \\
$\nu_{15}+\nu_{16}$ &	232	&	240	& \\
$\nu_{12}$          &	239	&	243	& 242	\\
$\nu_{24}$          &	253	&	253	& 264 	    \\
$\nu_{9}+\nu_{16}$  &	262	&	260	& 	\\
$\nu_{8}+\nu_{16}$  &	277	&	279	&    	\\
$4\nu_{16}$         &	281	&	280	&    	\\
$\nu_{15}+2\nu_{16}$ &	303	&	309	&    	\\
$\nu_{12}+\nu_{16}$ &	310	&	311	&  311	\\
$\nu_{24}+\nu_{16}$ &	323	&	322	&  	    \\
$2\nu_{15}$         &	324	&	330	&  336	\\
$\nu_{9}+2\nu_{16}$ &	332	&	340	&  	    \\
$\nu_{8}+2\nu_{16}$ &	347	&	348	&  	    \\
\bottomrule
    \end{tabular}
\end{table}

Finally, to perform the 8D intermolecular-plus-torsion vibrational computations, we first determined a potential-optimized DVR for every vibrational degree of freedom following the procedure described in Ref.~\citenum{DaAvMa21}. All remaining, i.e., constrained, degrees of freedom have been fixed at the values of the \pesfinal\ equilibrium geometry (cf. Table~\ref{tab:geom}).
The one-dimensional cuts of the three investigated HDNNPs and the 
QB16 potential along the intermolecular coordinates plotted in Fig.~4 show that the newly developed \pesfinal\ behaves well along all the intramolecular degrees of freedom consisting of the distance between the monomers and relative orientation (cf. Fig.~\ref{fig:fadcoor}). In contrast to the QB16 potential, which shows a low-energy oscillation for $\phi \approx 150^{\circ}$, no artificial cutoff of the primitive grid intervals has been found to be necessary for running the calculations. 

The obtained vibrational frequencies converged for the 8D($\mathcal{I}t$) intermolecular-plus-torsional representation are collected in Table~\ref{tab:8Dvib}. Overall the agreement for both QB16 and FAD-HDNNP with experiment is reasonably good with deviations of only a few ~\cm\ for most frequencies.

However, similarly to previous results discussed in Ref.~\citenum{DaAvMa21}, we can still observe a non-negligible blue shift in the problematic fundamental frequencies $\nu_8$ and $\nu_9$. With the present, extensively tested PESs, we can identify two possible origins and corresponding solutions for this shift. First, either it is a consequence of the constrained coordinates resulting in a too steep potential energy well at the equilibrium cut, which may be overcome by relaxing the constraint coordinates. Or, second, the number of the active vibrational degrees of freedom should be increased in the GENIUSH program. However, in the latter case the computational cost would increase rapidly in DVR, and for this reason, it will be necessary to use the more efficient FBR-Smolyak representation\cite{AvMa19,AvMa19b,DaAvMa22}. 
In comparison with VPT2 (cf. Sec.~\ref{sec:vpt2num}), it is interesting to note that VPT2 gets all fundamental frequencies (including $\nu_8$ and $\nu_9$) mostly correct (cf. Tab.~\ref{tab:frq_vpt2}). In Fig.~5 of Ref.~\citenum{NeSu20}, Nejad and Suhm contrasted the good performance of VPT2 ($\pm 5$~\cm\ with respect to experiment) against the very large deviations (15--40~\cm) of the fundamentally more complete VCI computations \cite{QuBo19}. Their observations motivated further theoretical and computational work, including an inquiry about the efficiency of normal coordinates for the intermolecular (low-frequency) vibrations for this system~\cite{DaAvMa21}.

\section{Summary \& Conclusions}
In this work we have examined the accuracy of machine-learned potential energy surfaces using the prototypical case of high-dimensional neural network potentials. Like many other modern machine learning potentials, HDNNPs have been primarily developed to transfer the accuracy of electronic structure methods to very large systems containing thousands of atoms, with the aim to perform large-scale molecular dynamics simulations. Consequently, the validation of the resulting multidimensional PESs is a difficult task. In this benchmark study, we have selected a system of moderate size, the formic acid dimer, for which harmonic vibrational frequencies at the CCSD(T) level of theory are accessible as a probe for assessing the quality of the PES along with a wealth of experimental data.

For the development of the full-dimensional HDNNP we have pursued a three-step procedure based on increasing reference data sets in the training process. First, a generally faithful representation of the PES without artifacts like artificial ``holes'' of overall good quality is generated. Second, this surface is then iteratively refined by adding further points to ensure that the harmonic frequencies of the PES reproduce within $\pm$10~\cm\ the harmonic frequencies of the CCSD(T) level of theory, which serves as our reference. Third, the representation of (lower-order) couplings is investigated by computing the VPT2 frequencies and the corresponding normal-coordinate quartic force constants in comparison with experimental data.

This carefully validated full-dimensional FAD-HDNNP surface has been used in (pilot) 8-dimensional, curvilinear, variational computations focusing on the low-energy intermolecular range. Further progress in the variational vibrational methodology is required for reaching a higher-energy spectral range range including a higher number of active vibrational degrees of freedom.

The finally obtained FAD-HDNNP potential energy surface shows a very high quality, which, in combination with the wealth of available, high-quality experimental data,\cite{NeSu20,GEORGES2004187,MatyRiehn,FumTai,MiPi1958,BeMi,BertieJohn,GanHal} we expect to be very useful for future developments in quantum dynamics and spectroscopic applications, which rely on robust and accurate PESs. 
\section{Conflicts of interest}
There are no conflicts to declare.
\begin{acknowledgments}
This project has been funded by the Deutsche Forschungsgemeinschaft (DFG, German Research Foundation) - 389479699/GRK2455. EM is grateful for a DFG Mercator Fellowship granted in the framework of this project.
AMSD and EM are grateful for the financial support of the Swiss National Science Foundation 
(PROMYS Grant, No.~IZ11Z0\_166525).
We thank Chen Qu and Joel Bowman for sharing with us their formic acid dimer PESs for providing their ab initio data set. Computing time has been available through the DFG project INST186/1294-1 FUGG (Project No. 405832858).
Discussions with Ricardo Mata, Rainer Oswald, Arman Nejad, Martin Suhm and Guntram Rauhut are gratefully acknowledged. 
\end{acknowledgments}

\bibliography{literature}
\end{document}


\title{Supporting Information to:High-dimensional neural network potentials for accurate vibrational frequencies: The formic acid dimer benchmark$^\dag$}

\author{Dilshana Shanavas Rasheeda$^{{\ast}1}$,Alberto Mart\'in Santa Dar\'ia\textit{$^{2}$}, Benjamin Schr\"oder$^{1}$,Edit M\'atyus$^{2}$, J\"org Behler}

\affiliation{\textit{$^{1}$~Universit\"at G\"ottingen, Institut f\"ur Physikalische Chemie, G\"ottingen, Germany
\\
$^{2}$~Eötvös Loránd University, Institute of Chemistry, Budapest, Hungary}}
\email{dilshana.rasheeda@chemie.uni-goettingen.de}

\maketitle

\onecolumngrid
\setcounter{table}{0}
\setcounter{figure}{0}
\setcounter{equation}{0}
\setcounter{section}{0}
\renewcommand{\thetable}{S\arabic{table}}%
\renewcommand{\thefigure}{S\arabic{figure}}%
\renewcommand{\theequation}{S\arabic{equation}}
\renewcommand{\thesection}{S-\Roman{section}}

\section{Construction of the HDNNPs}

The common settings for the construction of high-dimensional neural network potentials (HDNNPs) is given in Table \ref{tab:cons_nn}. For different iterations of the HDNNPs, different atom centred symmetry functions (ACSFs) were employed. ACSFs are used to represent the atomic environments and ensure translational and rotational invariance. For the construction of each of the iterations of the HDNPPs, additional structures are added (see main manuscript for details). As a result, the atomic environment also has to be redefined within each iteration. The parameters\cite{P2882} of the ACSFs used for constructing the various HDNNPs are given in Tables \ref{tab:ACSF-1},\ref{tab:ACSF-2},\ref{tab:ACSF-3} and \ref{tab:ACSF-ang}. The cutoff radius, as needed for the definition of the ACSFs,\citenum{P2882} is 14.901 Bohr for HDNNP1 and 15.0 Bohr for HDNNP2 as well as HDNNP3 i.e. the final FAD-HDNNP intended for production use. 

\begin{table}[h]
    \centering
    \caption{RuNNer settings for HDNNPs}
    \begin{tabular}{cc} \hline
       Keyword  & Settings \\ \hline
       nn\_type\_short  & 1 \\
       random\_number\_type &   5 \\
       global\_activation\_short & t t l \\
       cutoff\_type  &    1   \\
       use\_short\_nn   &       \\
       global\_hidden\_layers\_short &   2 \\
       scale\_symmetry\_functions & \\
       center\_symmetry\_functions & \\
     \hline  
    \end{tabular}
    \label{tab:cons_nn}
\end{table}

\begin{table*}[h]
     \centering
     \caption{Radial ACSF parameters $\eta$ for HDNNP1}
     \begin{tabular}{c l} \hline
     element pair & $\eta$[Bohr$^{-2}$] \\ \hline
      H-H    & 0, 0.003320, 0.007822, 0.014296, 0.024263, 0.040982, 0.072561, 0.144102 \\
      O-O    & 0, 0.002331, 0.005208, 0.008869, 0.013680, 0.020235, 0.029556, 0.043520   \\      C-C    & 0, 0.000964, 0.002013, 0.003161, 0.004425, 0.005824     \\
      H-C    & 0, 0.003763, 0.009087, 0.017202, 0.030743, 0.056242, 0.113944, 0.295433    \\
      O-C    & 0, 0.003648, 0.008752, 0.016415, 0.028926, 0.051756, 0.100815, 0.240433     \\
      H-O    & 0, 0.003910, 0.009520, 0.018245, 0.033218, 0.062638, 0.134133, 0.395239     \\ \hline
     \end{tabular}
     
     \label{tab:ACSF-1}
 \end{table*}
 
 \begin{table*}[h]
     \centering
     \caption{Radial ACSF parameters $\eta$ for HDNNP2 and HDNNP2a}
     \begin{tabular}{c l} \hline
     element pair & $\eta$[Bohr$^{-2}$] \\ \hline
      H-H    & 0, 0.004000, 0.009000, 0.016000, 0.028000, 0.049000, 0.094000, 0.215000 \\
      O-O    & 0, 0.003000, 0.006000, 0.010000, 0.015000, 0.022000, 0.032000, 0.048000   \\
      C-C    & 0, 0.003747, 0.009066, 0.017212, 0.030893, 0.056909     \\
      H-C    & 0, 0.004000, 0.009000, 0.018000, 0.031000, 0.056000, 0.114000, 0.296000    \\
      O-C    & 0, 0.004000, 0.009000, 0.017000, 0.029000, 0.052000, 0.101000, 0.241000     \\
      H-O    & 0, 0.004000, 0.010000, 0.019000, 0.033000, 0.063000, 0.134000, 0.395000    \\ \hline
     \end{tabular}
     
     \label{tab:ACSF-2}
 \end{table*}
 
 \begin{table*}[h]
     \centering
     \caption{Radial ACSF parameters $\eta$[Bohr$^{-2}$] for HDNNP3(FAD-HDNNP)}
     \begin{tabular}{c l} \hline
     element pair & $\eta$[Bohr$^{-2}$] \\ \hline
      H-H    &  0, 0.004000, 0.009000, 0.016000, 0.028000, 0.049000, 0.094000, 0.215000\\
      O-O    &  0, 0.003000, 0.006000, 0.010000, 0.015000, 0.022000, 0.032000, 0.048000  \\
      C-C    &  0, 0.003747, 0.009066, 0.017212, 0.030893, 0.056909    \\
      H-C    &  0, 0.004000, 0.009000, 0.018000, 0.031000, 0.056000, 0.114000, 0.296000    \\
      O-C    &  0, 0.004000, 0.009000, 0.017000, 0.029000, 0.052000, 0.101000, 0.241000    \\
      H-O    &  0, 0.004000, 0.010000, 0.019000, 0.035000, 0.067000, 0.149000, 0.486000    \\ \hline
     \end{tabular}
     
     \label{tab:ACSF-3}
 \end{table*}
 
 \begin{table}[h!]
     \centering
     \caption{Angular ACSF parameters $\eta$ for all element combinations and HDNNPs} 
     \begin{tabular}{c c c c } \hline
     No. & $\eta$[Bohr$^{-2}$] & $\zeta$    & $\lambda$  \\ \hline
       1    &  0.0 &  1.0 &  1.0  \\
       2    &  0.0 &  2.0 &  1.0  \\
       3    &  0.0 &  4.0 &  1.0  \\
       4    &  0.0 & 16.0 &  1.0  \\
       5    &  0.0 &  1.0 & -1.0  \\
       6    &  0.0 &  2.0 & -1.0  \\
       7    &  0.0 &  4.0 & -1.0  \\
       8    &  0.0 & 16.0 & -1.0  \\ \hline
     \end{tabular}
     
     \label{tab:ACSF-ang}
 \end{table}

\section{Quartic Force Constants}

The parameters of the quartic force field (QFF) for the formic acid dimer (FAD) were obtained with FAD-HDNNP by numerical differentiation. To this end, diagonalization of the (mass-weighted) numerical hessian yielded the harmonic frequencies $\omega_i$ and normal coordinate displacement vectors. Table~\ref{tab:minimum} provides the cartesian coordinates of the FAD-HDNNP minimum geometry. The cubic and quartic force constants $\phi_{ijk}$ and $\phi_{ijkl}$ (see manuscript and Ref.~\citenum{Clabo1988} for a definition) are calculated by standard finite difference formulas with up to 5 points per coordinate. A uniform value of 0.01 was chosen for the step size in terms of the dimensionless normal coordinates. Due to symmetry restrictions a large number of the $\phi_{ijk}$ and $\phi_{ijkl}$ vanish, i.e. only those are different from zero for which the direct product for the irreducible representations of the involved normal coordinates is totally symmetric. The resulting force constants of the QFF are provided in the formatted ASCII file \textsc{qff}. The Coriolis $\zeta^\alpha_{ij}$ needed to reproduce the VPT2 results provided in the main manuscript are also provided in an ASCII file named \textsc{zeta}. In these files only non-vanishing combinations of $\{i,j,k,l\}$ with $i\le j\le k\le l$ are quoted. Note that for standard VPT2 only the semi-diagonal QFF is relevant,\cite{Papousek1982} i.e. only those quartic force constants of type $\phi_{iiii}$ and $\phi_{iijj}$ contribute to the transition frequencies.

\begin{table}
     \centering
     \caption{\label{tab:minimum}Cartesian coordinates [\AA] of the FAD minimum structure obtained with FAD-HDNNP.} 
     \begin{tabular}{lD{.}{.}{10}p{0.1mm}D{.}{.}{10}D{.}{.}{10}}
    \hline
     Atom & x && y & z\\
   \hline
        H & 0.2631043230 && 0.0000000000 & 2.9819981733\\
        H &-0.2631043230 && 0.0000000000 &-2.9819981733\\
        H &-1.1172113912 && 0.0000000000 & 0.5090506679\\
        H & 1.1172113912 && 0.0000000000 &-0.5090506679\\
        O & 1.1554707969 && 0.0000000000 & 1.1679005999\\
        O &-1.1554707969 && 0.0000000000 &-1.1679005999\\
        O &-1.0747774476 && 0.0000000000 & 1.5017795982\\
        O & 1.0747774476 && 0.0000000000 &-1.5017795982\\
        C & 0.1767682702 && 0.0000000000 & 1.8926694202\\
        C &-0.1767682702 && 0.0000000000 &-1.8926694202\\
    \hline
     \end{tabular}
\end{table}

\newpage

 \bibliography{literature}